\documentclass[%
 reprint,
superscriptaddress,
nofootinbib,
 amsmath,amssymb,
 aps,
pra,
]{revtex4-2}

\usepackage{graphicx}% Include figure files
\usepackage{dcolumn}% Align table columns on decimal point
\usepackage{bm}% bold math
\usepackage{siunitx}

\usepackage[caption=false, position=top]{subfig}
\usepackage{hyperref}
\hypersetup{
    colorlinks = true,
    linkcolor = blue
    }
\usepackage[
        noabbrev,
        capitalise,
        nameinlink,
    ]
    {cleveref}

\usepackage{amsmath}
\usepackage{amssymb}
\usepackage{bm}
\usepackage{bbm}

\begin{document}

%\preprint{APS/123-QED}

\title{Detrimental non-Markovian errors for surface code memory}

\author{John F Kam}\thanks{john.kam@monash.edu}\thanks{Author to whom any correspondence should be addressed.}
\affiliation{School of Physics and Astronomy, Monash University, Clayton, VIC 3168, Australia}
\affiliation{Quantum Systems, Data61, CSIRO, Clayton, VIC 3168, Australia}
\author{Spiro Gicev}
\affiliation{School of Physics, The University of Melbourne, Parkville, VIC 3052, Australia}
\author{Kavan Modi}%\thanks{Kavan.Modi@monash.edu}
\affiliation{Science, Mathematics and Technology Cluster, Singapore University of Technology and Design, 8 Somapah Road, 487372 Singapore}
\affiliation{School of Physics and Astronomy, Monash University, Clayton, VIC 3168, Australia}
\author{Angus Southwell}
\affiliation{School of Physics and Astronomy, Monash University, Clayton, VIC 3168, Australia}
\author{Muhammad Usman}%\thanks{Muhammad.Usman@data61.csiro.au}
\affiliation{Quantum Systems, Data61, CSIRO, Clayton, VIC 3168, Australia}
\affiliation{School of Physics, The University of Melbourne, Parkville, VIC 3052, Australia}
\affiliation{School of Physics and Astronomy, Monash University, Clayton, VIC 3168, Australia}

\date{\today}% It is always \today, today,

\begin{abstract}
The realization of fault-tolerant quantum computers hinges on effective quantum error correction protocols, whose performance significantly relies on the nature of the underlying noise. In this work, we directly study the structure of non-Markovian correlated errors and their impact on surface code memory performance. Specifically, we compare surface code performance under non-Markovian noise and independent circuit-level noise, while keeping marginal error rates constant. Our analysis shows that while not all temporally correlated structures are detrimental, certain structures, particularly multi-time “streaky” correlations affecting syndrome qubits and two-qubit gates, can severely degrade logical error rate scaling. Furthermore, we discuss our results in the context of recent quantum error correction experiments on physical devices. 
These findings underscore the importance of understanding and mitigating non-Markovian noise toward achieving practical, fault-tolerant quantum computing.
\end{abstract}

\maketitle

%\tableofcontents

%##############################################################
%
% SECTION: INTRODUCTION
%
% ##############################################################

\section{Introduction}
\begin{figure*}
    \centering
    \includegraphics[width=.8\textwidth]{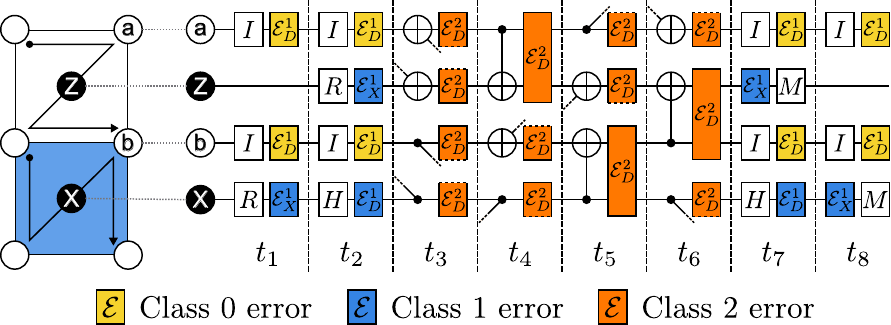}
    \caption{\label{fig:singleround}A single round of the rotated surface code circuit under circuit-level noise. Class 0 errors (colored in yellow), or idling errors, affect only data qubits. Class 1 errors (colored in blue), or SPAM errors, affect only syndrome qubits. Class 2 errors (colored in orange), or control errors, which model noisy two-qubit entangling gates, affect both data and syndrome qubits. $\mathcal{E}^1_D$ is the single-qubit depolarizing error channel, $\mathcal{E}^1_X$ is the single-qubit bit-flip channel, and $\mathcal{E}^2_D$ is the two-qubit depolarizing error channel.
    }
\end{figure*}
The full potential of quantum computers is expected to emerge once fault-tolerant quantum computers have been realized. The main pathway to reach fault tolerance is quantum error correction (QEC) \cite{p_w_shor_fault-tolerant_1996, aharonov_fault-tolerant_1997, knill_resilient_1998}, which entails redundantly encoding the quantum information of a few qubits onto the logical subspace of many physical qubits. However, the efficacy of such protocols heavily depends on the underlying noise characteristics affecting the physical system. Crucially, complex, correlated noise can break critical assumptions that quantum error-correcting codes rely on to suppress logical errors \cite{novais_surface_2013, fowler_quantifying_2014, clader_impact_2021, sriram_non-uniform_2024}. Experimentally, noise affecting QEC protocols shows signs of a rich underlying structure \cite{acharya_suppressing_2023, bluvstein_logical_2024,acharya_quantum_2024}, leading to numerous studies focused on mitigating its adverse effects. Beyond spatial inhomogeneity, quantum devices may experience biased noise \cite{tuckett_ultrahigh_2018, tuckett_tailoring_2019,tuckett_fault-tolerant_2020}, coherent errors \cite{gutierrez_errors_2016, greenbaum_modeling_2017}, crosstalk \cite{murali_software_2020, sarovar_detecting_2020, parrado-rodriguez_crosstalk_2021, tripathi_suppression_2022}, noise drift \cite{proctor_detecting_2020, wang_dgr_2024}, leakage \cite{chen_measuring_2016, wood_quantification_2018, mcewen_removing_2021, miao_overcoming_2023}, cosmic ray bursts \cite{martinis_saving_2021,y_suzuki_q3de_2022,wilen_correlated_2021,siegel_adaptive_2023}, and generally non-Markovian noise \cite{white_demonstration_2020,PhysRevApplied.15.014023,white_non-markovian_2022,white2022manybodymanytimephysics,white_unifying_2023}. Non-Markovian or temporally correlated noise, in particular, has not been extensively studied in the context of surface codes. 

Correlated noise---spatial, temporal, or both---can be detrimental to QEC protocols for several reasons. Firstly, it can increase the average physical error rate beyond threshold levels. A recent study found that rare regions of elevated error rates can dominate the logical behavior of topological quantum memory \cite{sriram_non-uniform_2024}. Moreover, the structure of correlated errors can interact non-trivially with topological-based logical operations. Recent experiments on superconducting qubits demonstrated exponential suppression of logical error rates up to distance 7 \cite{acharya_quantum_2024}. However, a logical-error-per-cycle floor around $10^{-10}$ was observed for high-distance repetition codes. This noise floor was attributed to rare correlated error events, including spatially correlated large-area errors and temporally correlated noisy single-detector events. As these effects are likely to eventually dominate logical error rates at lower physical error rates and larger code distances, understanding and mitigating correlated errors becomes increasingly crucial.

In this paper, we numerically study the performance impact of temporally correlated errors on surface codes---a prominent family of quantum error-correcting codes that are key to many strategies for achieving practical, fault-tolerant quantum computing \cite{kitaev_quantum_1997, kitaev_fault-tolerant_2003, bravyi_quantum_1998, fowler_towards_2012}. Key to our analysis is comparing surface code performance under correlated noise versus independent noise while keeping marginal error rates of each qubit constant, thereby isolating the impact of the structure of noise rather than the overall error rate. %We test various temporally correlated error structures derived from time correlations in circuit-level noise. 
Our models explore a range of temporally correlated error patterns, from pairwise to multi-time correlations, and polynomially decaying to exponentially decaying correlations. We find that certain temporally correlated noise structures are catastrophic to logical error rate scaling in surface code performance. Specifically, multi-time streaks of correlated errors affecting syndrome qubits, or two-qubit entangling gates, can result in strongly slower-than-exponential logical error rate suppression in quantum memory experiments. These findings highlight the need for deeper understanding and mitigation of such correlated errors in the pursuit to build a fault-tolerant quantum computer.

The paper is organized as follows: In \cref{sec:background}, we provide an overview of key concepts related to surface codes, circuit-level noise, detectors and detector error models. We then motivate temporally correlated noise models for surface codes based on the current experimental landscape. In \cref{sec:fullcircuitlevel}, we present our temporally correlated circuit-level noise models and numerical simulation results for two-time and multi-time correlated error models with quadratically decaying correlations. \Cref{sec:errorclasses} analyzes temporal correlations by error class to identify which types of correlated errors most significantly impact surface code memory. Finally, \cref{sec:discussion} offers a perspective on how temporally correlated error structures can negatively impact 2D topological memory performance, potential strategies to mitigate these effects, and concludes the paper.

% ##############################################################
%
% SECTION: BACKGROUND
%
% ##############################################################
\section{Background}\label{sec:background}
Before presenting our analyses and results, we briefly introduce surface codes, circuit-level noise, and temporally correlated noise. This section covers error classes, detectors, and detector error models, providing the necessary context to understand the significance of temporally correlated noise in surface code experiments.

\subsection{Surface codes and circuit-level noise models}
For our purposes, a distance $d$ surface code is a $[[d^2, 1, d]]$ Calderbank-Shor-Steane (CSS) code that encodes a single logical qubit across $d^2$ data qubits and $d^2-1$ syndrome qubits embedded on a two-dimensional grid \cite{bombin_optimal_2007}. Each data qubit interacts locally with up to four syndrome qubits, where half of the syndrome qubits are used to perform projective measurements in the $Z$ basis, and the other half are used to perform projective measurements in the $X$ basis. In this paper, while we specifically study the rotated XZ surface code, our analyses of temporally correlated errors should extend to other surface code variations including the XY surface code \cite{tuckett_tailoring_2019}, and the XZZX surface code \cite{bonilla_ataides_xzzx_2021}.

Under a sufficiently sparse distribution of physical errors, a decoding algorithm is able to infer the location and types of errors from the syndrome measurement outcomes, and thereby return a recovery operation to restore the logically encoded state. In a single error correction round, i.e., applying the full syndrome extraction circuit, the surface code is guaranteed to correct up to weight (number of simultaneous errors) $L = \lfloor(d-1)/2\rfloor$ errors on data qubits. Under noisy syndrome extraction, syndrome qubit errors are inferred from multiple rounds of syndrome measurements. Assuming an uncorrelated error model, higher distance codes will require more rounds, on the order of $\mathcal{O}(d)$, to sufficiently suppress syndrome qubit errors. This has the disadvantage however of exposing the surface code to temporally correlated errors.

Historically, the surface code and its threshold were first analyzed under phenomenological noise models \cite{dennis_topological_2002,wang_confinement-higgs_2003}---models that consist only of data qubit errors outside syndrome extraction cycles, and classical measurement errors. More recently, with the advent of physical quantum devices with sizes sufficient to support QEC experiments \cite{acharya_suppressing_2023, bluvstein_logical_2024,acharya_quantum_2024}, it has become more common to model noise on the circuit level. Circuit-based noise models attempt to model physical device noise more realistically by modeling errors on the level of noisy gate operations on physical qubits. Errors can occur anywhere throughout the syndrome extraction circuit, and the details of the noise model can be fine-tuned to specific physical scenarios. 

\Cref{fig:singleround} illustrates circuit-level noise for a single surface code round on a four qubit subset of two data qubits and two syndrome qubits. Each physical gate is followed by an error channel. Errors are divided into three classes \cite{fowler_towards_2012}: idling or memory errors on data qubits, or Class 0 errors, state preparation or measurement errors (can be quantum or classical) on syndrome qubits, or Class 1 errors, and two-qubit entangling gate errors on data and syndrome qubit pairs, or Class 2 errors. \Cref{fig:singleround} shows three different error channels: the single-qubit bit-flip channel $\mathcal{E}^1_X$, the single-qubit depolarizing channel $\mathcal{E}^1_D$, and the two-qubit depolarizing channel $\mathcal{E}^2_D$, defined as
\begin{subequations}\label{eqn:channels}
    \begin{align}
    &\mathcal{E}^1_X(p{,\rho}) = (1-p)\rho + pX\rho X\label{eqn:channelX} \\
    &\mathcal{E}^1_D(p{,\rho}) = (1-p)\rho + \frac{p}{3}(X\rho X + Y\rho Y + Z\rho Z)\label{eqn:channelD1} \\
    &\mathcal{E}^2_D(p{,\rho}) = (1-p)\rho + \frac{p}{15}\sum_{P\in\{I,X,Y,Z\}^{\otimes 2}\setminus \{II\}}P\rho P\label{eqn:channelD2}.
    \end{align}
\end{subequations}
In standard circuit-level noise $p$ is identical for each channel. Note that in some variations of standard circuit-level noise, Class 0 errors are concatenated to a single identity gate followed by a depolarizing error channel at the beginning of the round, which will be the convention adopted in this paper.

\subsection{Detector error models}
To help build a working understanding of how the surface code detects and corrects errors on the circuit level, it is useful to define \textit{detectors}, \textit{detection events}, and \textit{detector error models} \cite{gidney_stim_2021, mcewen_relaxing_2023, derks_designing_2024}. 

A detector $D_{s, t}$ is an operation on a set of measurement outcomes, in this case from syndrome measurements $M_{s, t}$ with a deterministic outcome in the absence of errors. Here $s=(x, y)$ are the spatial coordinates, corresponding to the coordinates of syndrome qubits, and $t$ are the time coordinates. For an $N$ round ($t\in\{1, 2, ..., N+1\}$) surface code logical memory $Z$ (computational basis) experiment with $m$ syndrome qubits, we define each detector as
\begin{equation}
    D_{s,t} = M_{s, t-1} \oplus M_{s, t},
\end{equation}
where $\oplus$ represents modulo 2 addition. Detector $D_{s, t}$ is effectively a parity operation on $M_{s, t}$ at times $t-1$ and $t$. Note that $M_{s, t}$ is only explicitly measured at times $t\in \{1, 2, ..., N\}$. $M_{s, 0}$ is inferred directly from the data qubit state preparation step, and $M_{s, N+1}$ is inferred from the final data qubit measurement outcomes, defined as $M_{s, N+1} = \bigoplus_{u\in G(s)}M_{u, N+1}$, where $M_{u, N+1}$ is the final measurement outcome of a neighboring data qubit at spatial coordinates $u\in G(s)$. Furthermore, for a logical $Z$ basis memory experiment, detection events are not defined at $X$ syndromes at $t=1$ and $t=n+1$ (and vice versa for $X$ basis memory experiments), resulting in $\frac{m}{2} + (N-1)m + \frac{m}{2} = N\times m$ total number of detectors.

In the complete absence of errors, each detector deterministically outputs $D_{s,t}=0$. In the presence of errors, a measurement outcome at syndrome qubit $s$ can be flipped between rounds $t-1$ and $t$, resulting in a detector outcome of $D_{s,t}=1$, called a \textit{detection event}. The locations of detection events reveal details on which errors are likely to have occurred. For instance, a Pauli $X$ or $Z$ error on a data qubit at the beginning of an error correction round will trigger a pair of adjacent detection events at different spatial coordinates and the same time coordinates. Such errors are labeled \textit{spacelike} errors. On the other hand, an $X$ error on a syndrome qubit right before measurement will trigger a pair of detection events at the same spatial coordinates but different time coordinates. These errors are labeled \textit{timelike} errors. There also exist \textit{spacetimelike} errors, or \textit{hook} errors, that describe errors in the middle of the syndrome extraction circuit, which can be caused by faulty two-qubit entangling gate operations.

We are now able to define a \textit{detector error model}, which is simply a dictionary containing every possible physical error mechanism in a Clifford circuit, the set of detection events it triggers, and its probability of occurring as prescribed by a noise model. The detector error model provides a convenient framework for interfacing circuit-based noise models with decoding algorithms. For instance, the detector error model for \cref{fig:singleround} can be represented graphically, where detectors are vertices and error mechanisms are weighted edges - hence allowing for straightforward implementation of graph-based decoding algorithms such as minimum weight perfect matching (MWPM) \cite{wang_confinement-higgs_2003,fowler_minimum_2014}.

In the context of correlated noise models, it is convenient to define \textit{stringlike} errors, which are chains of multiple errors that combine in such a way that they only trigger up to two detection events: one at each end of the chain, one at one end of a chain and the other end terminating at a boundary, or both ends terminating at a boundary. A stringlike error can be purely spatial (a line of data qubit errors in a single round), purely temporal (a line of syndrome qubit errors over multiple rounds), or a mixture of both. Stringlike errors are among the most difficult error structure to correct, as a single spatial stringlike error of length $L = \lceil(d+1)/2\rceil$ is enough to cause a logical error in a distance $d$ surface code. Under an independent, identically distributed (i.i.d.) noise model, the probability of a stringlike error of length $L$ decays exponentially with $L$ in the small $p$ limit. Under certain correlated noise models however, the probability distribution of $L$ length stringlike errors can exhibit power law scaling (i.e., heavy-tails), which can be extremely detrimental to achieving fault tolerance, even if the average physical error rate is below threshold \cite{fowler_quantifying_2014,clader_impact_2021}. Next, we delve further into describing temporally correlated noise.

\subsection{Motivating temporally correlated noise models}
\begin{figure*}[t]
    \centering
    \subfloat[\label{fig:corr_error_circuit}]
    {\includegraphics[width=\textwidth]{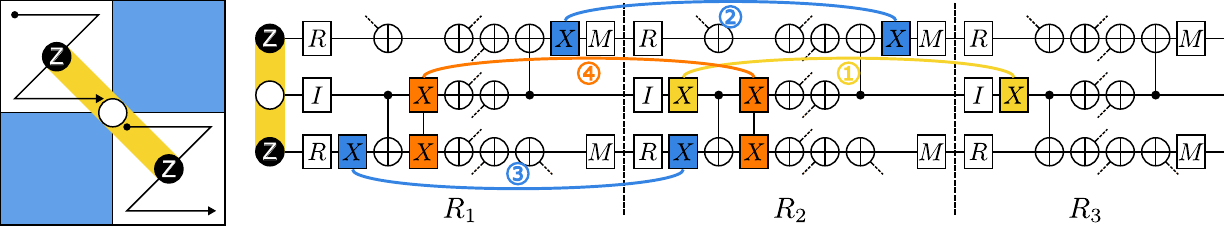}}
    
    \subfloat[\label{fig:corr_memory}]
    {\includegraphics[width=.23\textwidth]{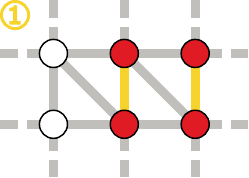}}
    \hspace{3mm}
    \subfloat[\label{fig:corr_measure}]
    {\includegraphics[width=.23\textwidth]{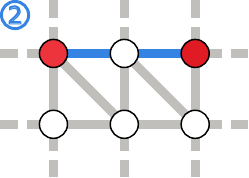}}
    \hspace{3mm}
    \subfloat[\label{fig:corr_reset}]
    {\includegraphics[width=.23\textwidth]{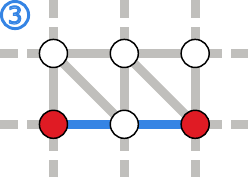}}
    \hspace{3mm}
    \subfloat[\label{fig:corr_cx}]
    {\includegraphics[width=.23\textwidth]{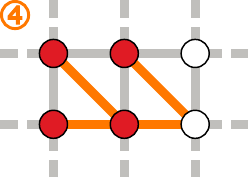}}
    \caption{\label{fig:correlatederrormechanisms}Various two-time correlated errors and their corresponding syndromes. \protect\subref{fig:corr_error_circuit} displays three rounds of a surface code syndrome measurement circuit for a data qubit and two $Z$ syndrome qubits. We illustrate several cases of two-time bit-flip error channels (spanning two rounds), including a correlated idling error (Class 0), a correlated measurement error (Class 1), a correlated reset error (Class 1), and a correlated two-qubit CNOT error (Class 2). \protect\subref{fig:corr_memory} the error syndrome due to a pair of idle $X$ errors at $R_2, R_3$ on the data qubit, this creates two spacelike edges resulting in four detection events (red vertices, corresponding to a change in syndrome measurement outcome compared to the previous round). \protect\subref{fig:corr_measure} the error syndrome due to a pair of measurement $X$ errors at $R_1, R_2$ on the top syndrome qubit, this creates two timelike edges resulting in two detection events. \protect\subref{fig:corr_reset} the error syndrome due to a pair of reset $X$ errors at $R_1, R_2$ on the top syndrome qubit, this creates two timelike edges resulting in two detection events. \protect\subref{fig:corr_cx} the error syndrome due to a pair of CNOT $XX$ errors at $R_1, R_2$, this creates two spacetimelike edges and two timelike edges, resulting in four detection events.}
\end{figure*}
When surveying QEC literature, one generally comes across three types of noise correlations. The first type describes correlations between types of errors, such as $X$/$Z$ error correlations. For example, in the quantum depolarizing channel, $X$ (bit-flip) and $Z$ (phase-flip) errors are correlated due to the presence of $Y$ (combined flip) errors. Several studies have sought to improve the performance of decoding algorithms by taking into account error type correlations \cite{duclos-cianci_fast_2010, wootton_high_2012, fowler_optimal_2013, n_delfosse_decoding_2014, criger_multi-path_2018}. The second type of correlated noise describes correlations in space. A body of research has been dedicated to modeling, characterizing, and addressing spatially correlated errors in surface codes \cite{fowler_quantifying_2014, nickerson_analysing_2019,harper_efficient_2020,chubb_statistical_2021,harper_learning_2023}. Ref.~\cite{fowler_quantifying_2014} measures the impact of spatially correlated many-qubit errors, showing at least moderate exponential suppression of large-area errors is required for tolerable qubit overheads. Ref.~\cite{nickerson_analysing_2019} develops methods to characterize and mitigate spatial stringlike correlated errors via adaptive decoding algorithms. Many studies have also focused on addressing specific physical error processes that are known to exhibit significant spatial correlations in surface codes. These notably include studies on leakage errors \cite{battistel_hardware-efficient_2021,miao_overcoming_2023} and cosmic ray bursts \cite{y_suzuki_q3de_2022,siegel_adaptive_2023,mcewen_resisting_2024}. Ref.~\cite{miao_overcoming_2023} demonstrates leakage mitigation techniques for a distance 3 surface code experiment on superconducting qubits.

The third type of correlated noise, and perhaps the least studied in the context of QEC, describes correlations in time. Many of the physical processes mentioned previously that give rise to spatially correlated noise similarly give rise to temporally correlated noise, in addition to factors such as fabrication defects \cite{auger_fault-tolerance_2017,strikis_quantum_2023,siegel_adaptive_2023} and noise drift \cite{proctor_detecting_2020, wang_dgr_2024}. More generally, non-Markovian noise is a typical feature of quantum systems interacting with a bath environment, where repeated system-bath interactions lead to temporally correlated noise. A body of work is dedicated to characterizing this noise in physical devices \cite{white_demonstration_2020,milz_quantum_2021,PhysRevApplied.15.014023,white2022manybodymanytimephysics,white_unifying_2023}. Ref.~\cite{tanggara_strategic_2024} provides an operational quantum dynamics-based framework for tailoring quantum error-correcting codes to non-Markovian noise. A family of error-correcting codes called single-shot codes, which can effectively decode syndrome measurement errors in a single round, are believed to be generally robust against time-correlated errors \cite{bombin_single-shot_2015, bombin_resilience_2016}. Such codes, however, impose additional requirements on qubit connectivity, and no two-dimensional topological code can be single-shot. Furthermore, it is worth noting the numerous studies addressing other features of complex noise, that often intersect with correlated noise, such as coherent noise \cite{jain_improved_2023} and biased noise \cite{tuckett_ultrahigh_2018,tuckett_tailoring_2019}.

To capture the complex and diverse physical processes underlying temporally correlated noise, our noise models should be flexible while remaining grounded in experimental realities. To help achieve this, we model temporal correlations at the circuit level, specifically between gates across multiple rounds of syndrome extraction. The dominant physical process driving correlated errors can vary throughout the syndrome extraction circuit, depending on the gate and where it is located. For instance, in superconducting platforms where syndrome extraction cycles are $\sim$1 \si{\micro\second} long, the combined measurement and reset step on syndrome qubits can last upwards of 40\% of the cycle duration \cite{acharya_quantum_2024}. Errors on data qubits will be dominated by free system-bath evolution dynamics (assuming no additional control techniques such as dynamical decoupling). On the other hand, correlated errors on two-qubit entangling gates may stem from more systemic control issues related to cross-talk, frequency collisions, or calibration drift \cite{sheldon_procedure_2016}. While correlations can certainly occur between different error classes, we focus our study on correlations between gate-based errors of similar physical origin.

\Cref{fig:correlatederrormechanisms} shows three surface code rounds and several examples of the kind of temporally correlated errors we explore throughout this paper. We illustrate four different two-time correlated bit-flip errors (but can be any Pauli error in principle),  including a correlated idling error (Class 0), correlated measurement error (Class 1), correlated reset error (Class 1), and correlated CNOT error (Class 2). The correlated errors act on the same qubit or pair of qubits separated by a single round. Furthermore, we illustrate which detectors each correlated error triggers, where red vertices correspond to detection events, and colored edges to error mechanisms. We note a difference between time correlated errors on data qubits and time correlated errors on syndrome qubits. The two-time correlated measurement or reset errors combine to form a weight-2 temporal stringlike error, whereas the two-time correlated idling error form two weight-one spacelike errors. 

Finally, after examining the types and locations of correlated errors---what and where errors are correlated---we address \textit{how} errors are correlated over time. Specifically, we explore the temporal extent of these correlations---how far apart errors are correlated---and the number of errors involved, e.g., two-time versus multi-time correlations. While the specifics depend on the underlying physical processes, temporal correlations are expected to weaken as time separation increases. For instance, elevated error rates in superconducting qubits caused by cosmic ray bursts typically decay exponentially over multiple surface code rounds \cite{mcewen_resolving_2022}. Similarly, temporal correlations may decay polynomially, as is characteristic of $1/f$ noise \cite{white_demonstration_2020}. We maintain a flexible approach by modeling a range of two-time and multi-time correlations with varying strengths.

\section{Surface code performance under temporally correlated circuit-level noise}\label{sec:fullcircuitlevel}
\subsection{Noise models}\label{sec:noisemodeldetails}
\begin{figure}
    \centering
    \includegraphics[width=\columnwidth]{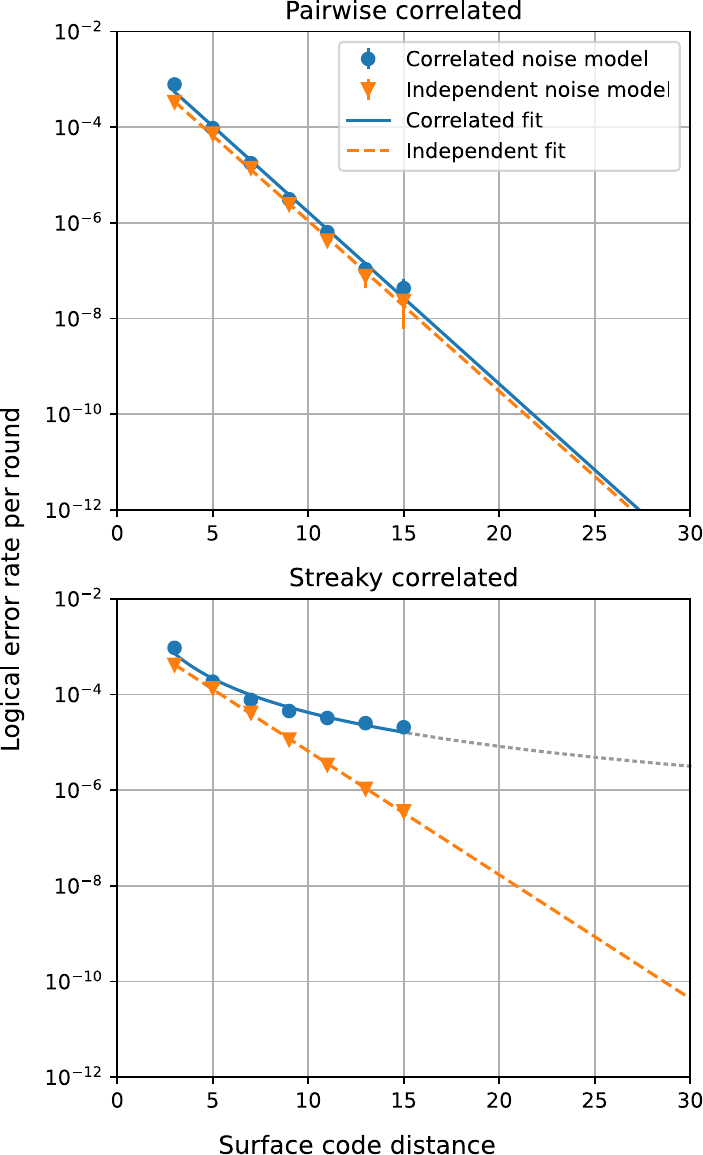}
    \caption{\label{fig:experiment3}Logical error rate per round versus surface code distance under temporally correlated circuit-level noise models. Error bars correspond to 95\% confidence intervals (up to marker size). Memory experiments are $2d$ rounds long and use minimum weight perfect matching to decode logical errors. Under the pairwise correlated noise model (top), error pairs occur with probability $Aq/\Delta t^2$, where $\Delta t=|t_1-t_2|$ is the time separation in rounds. Under the streaky correlated noise model (bottom), error streaks occur with probability $Aq/t^2$, where $t$ is the streak length in rounds. In both models, $A_\text{Class 0} = A_\text{Class 1} = 2A_\text{Class 2}$. In each plot, we compare performance under correlated noise model with performance under the marginalized independent model, where errors are temporally uncorrelated, but have the same marginal error rates as the correlated model.}
\end{figure}
\begin{figure*}
    \centering
    \includegraphics[width=0.8\textwidth]{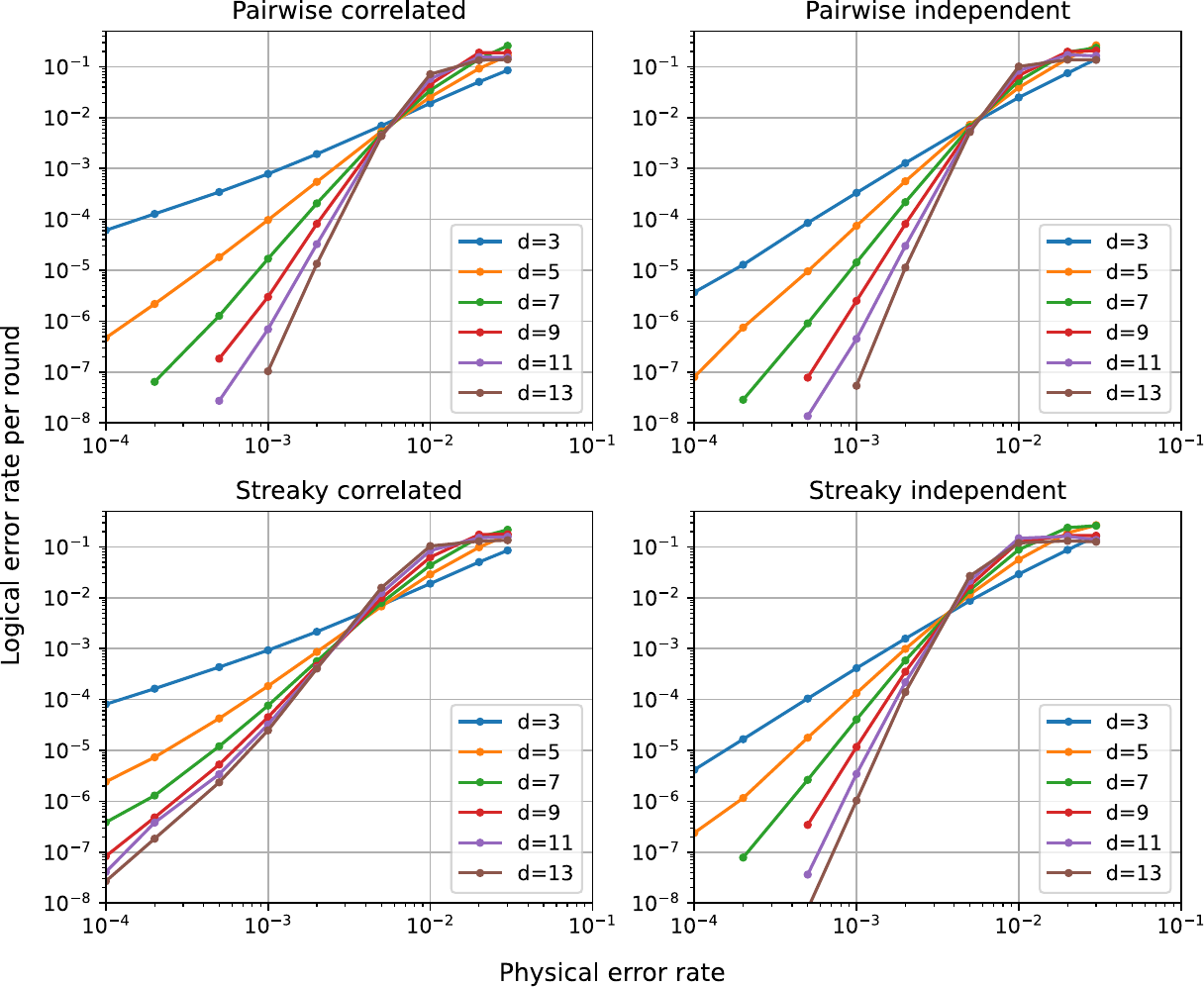}
    \caption{\label{fig:threshold}
    Threshold plots under temporally correlated circuit-level noise (pairwise and streaky) and marginalized independent noise. The point at which trendlines intersect, known as the accuracy threshold, represents the physical error rate below which increasing the code distance reduces the logical error rate. For both the pairwise correlated and marginalized independent model, we observe similar thresholds of approximately 0.6\%. Additionally, both models exhibit comparable scaling behavior between logical and physical error rates, as previously illustrated in \cref{fig:experiment3}. In contrast, for the streaky correlated and marginalized independent model, with respective \textit{apparent} thresholds of 0.3\% and 0.4\%, we observe slower-than-exponential suppression of logical error rates in the streaky correlated model up to distance 13.}
\end{figure*}
To isolate the impact of the structure of temporally correlated errors on surface code performance, we compare surface code performance under a temporally correlated model versus a temporally uncorrelated model while fixing the marginal error rates. This means that for each qubit at each timestep, an experimenter would measure the same error rate under both the correlated and uncorrelated models. Note that our temporally uncorrelated noise models can contain spatial correlations due to errors on two-qubit entangling gates. For details on calculating marginal errors rates from correlated error models, see \cref{sec:marginalization}. We only consider correlations within an error class; for example, Class 0 idling errors are only correlated with other Class 0 idling errors. 

We consider two types of temporally correlated error structures. The first structure involves \textit{pairwise} correlations, which introduce two-time correlated errors on the same qubit (or same pair of qubits in the case of Class 2 errors) at two different rounds. The two-time error channel depends on the class; for correlated Class 0 errors, a qubit experiences a two-time depolarizing channel at times $t, t'$ with errors denoted by $I_tX_{t'}, I_tY_{t'}, I_tZ_{t'}, ..., Z_tZ_{t'}$ each occurring with equal $1/15$ probability (this is the two-time equivalent to the two-qubit depolarizing channel shown in \cref{eqn:channelD2}). This model can be viewed as the temporal analog of the spatially correlated nonlocal two-qubit error model in Ref.~\cite{fowler_quantifying_2014}. For correlated Class 1 errors, a qubit experiences a two-time $X_tX_{t'}$ error, and for correlated Class 2 errors, a pair of qubits experiences a two-time, two-qubit depolarizing error after a CNOT operation.
Every round pair experiences a probability of a correlated pairwise error, equating to $\binom{N}{2}=N(N-1)/2$ probability rolls per qubit (since each qubit independently experiences temporally correlated errors), where $N$ is the number of rounds. The probability of any pair of errors decays polynomially or exponentially with round separation. In the polynomial decay model, a pair of errors occur with probability $Aq/\Delta t^n$, where $\Delta t=|t_1-t_2|$ is the distance in rounds, $q$ is the characteristic correlated error parameter and $A, n$ are model parameters. Similarly, in the exponential decay model, a pair of errors occur with probability $Aq/n^{\Delta t}$.

The second structure of temporally correlated errors is the \textit{streaky} model. In the streaky model, a qubit or pair of qubits experience an event which causes maximal mixing over a length of rounds $t$. For instance, in the case of correlated Class 0 errors, a data qubit becomes maximally depolarized, experiencing an equal probability of an $I, X, Y,$ or $Z$ error, during the idling portion of each round for $t+1$ rounds. For Class 1 errors, a syndrome qubits experiences an equal probability of an $I$ or $X$ error, i.e., a 0.5 probability of experiencing an $X$ error, and for Class 2 errors, a data-syndrome qubit pair experiences two-qubit maximally depolarizing noise after each CNOT operation. Similar to the pairwise model, a streak has a probability of occurring between every pair of rounds, and the length of rounds of maximal mixing decays polynomially or exponentially with $t$. Note physical error rates may be time inhomogeneous between rounds in either of these models (middle rounds will experience the highest error rate), however, marginalization ensures the independent model will be identically inhomogeneous to the correlated model.

\subsection{Numerical simulations}
We perform numerical simulations of surface code memory using Stim \cite{gidney_stim_2021}, a tool for simulating quantum stabilizer circuits. Since Stim does not natively support Monte Carlo sampling of temporally correlated errors, we develop custom sampling methods to introduce this type of noise. Specifically, we utilize the \texttt{stim.FlipSimulator} class, which enables dynamic injection of errors, in addition to executing stabilizer circuits and uncorrelated error sampling. At each timestep, \texttt{FlipSimulator} takes as input either a circuit layer operation, or a configuration of Pauli errors, represented by three binary vectors corresponding to the locations of $X$, $Y$, or $Z$ errors.

To sample correlated errors, we first generate an error matrix, or \textit{error mask}, $\bm{M} = M_{ij}$, where $i$ indexes spatial locations (qubits) and $j$ indexes timesteps. Each element $M_{ij}$ indicates whether an error channel has occurred on qubit $i$ at timestep $j$. The columns $\bm{M} = M_{i0}, M_{i1}, \ldots$ generate the binary vectors specifying the location of Pauli errors, which are input to \texttt{FlipSimulator} at each timestep. The specific Pauli error applied depends on the error channel at location $i$ and time $j$. For instance, if an error follows an identity gate, a random $X$, $Y$, or $Z$ error is chosen with equal probability, consistent with a depolarizing channel for circuit-level noise (\cref{fig:singleround}).

Elements of $\bm{M}$ are populated according to a predefined temporally correlated noise model. For example, in the Class 1 temporally correlated pairwise and streaky model, we first construct a probability matrix $\bm{R} = R_{ik}$, where each element $R_{ik}$ gives the probability of a correlated error affecting qubit $i$ at times $t_1$ and $t_2$, denoted by $k$. This probability is calculated according to the correlation decay model, either polynomial ($Aq/\Delta t^n$) or exponential ($Aq/n^{\Delta t}$), as defined in \cref{sec:noisemodeldetails}. Sampling $\bm{R}$ yields a binary outcome matrix $\bm{O} = O_{ik}$, indicating when and where correlated error events have occurred. These events in $\bm{O}$ are then mapped to individual errors in $\bm{M}$ via a transformation matrix $\bm{T} = T_{kj}$, which yields error mask elements $M_{ij}$ as
\begin{equation}\label{eqn:transform1}
    M_{ij} = \bigoplus_k O_{ik}T_{kj}.
\end{equation}
The form of $\bm{T}$ depends on the correlation structure. In the pairwise model, matrix elements $T_{kj}$ are one if $j \in \{t_1, t_2\}$; in the streaky model, $T_{kj}$ elements are one if $j \in [t_1, t_2]$, and zero otherwise. This procedure generalizes for other error classes, though specifics vary for two-qubit errors and depolarizing noise due to error composition properties. For Class 1 correlated errors, we apply a rescaling factor of $4/3$ to $\bm{R}$, and modify \cref{eqn:transform1} to
\begin{equation}\label{eqn:transform2}
    M_{ij} = \bigvee_k O_{ik}T_{kj},
\end{equation}
where $\bigvee$ is the logical OR operation. Full details are contained in the source code for this project \cite{github_2024}.

To decode errors, we employ minimum weight perfect matching via PyMatching \cite{higgott_pymatching_2022}, inputting the same marginalized independent detector error model for both sets of syndrome data. Each memory experiment is $2d$ rounds long (or two $d\times d \times d$ blocks). This choice is primarily practical, balancing computational constraints with the need to capture temporal correlations in sufficiently long memory experiments. In realistic fault-tolerant quantum computations, logical qubits often idle for many $d$ rounds, and these correlations are expected to persist or may become even more pronounced in longer experiments. We implement polynomially decaying correlations for both the pairwise and streaky correlated model, parameterized by $A_{\text{Class 0}} = A_{\text{Class 1}} = 2A_{\text{Class 2}} = 1$, and $n=2$.  We set the characteristic correlated error parameter $q=1\times 10^{-3}$.
%, or approximately a tenth of the threshold under standard circuit-level noise. 

\subsection{Results}\label{sec:experiment3}
\Cref{fig:experiment3} plots the logical error per round versus surface code distance for pairwise and streaky temporally correlated error models. Each experiment is run with 10 million shots, where error bars represent 95\% confidence intervals. We simulate memory experiments up to surface code distance $d=15$, and apply an exponential or power law fit, depending on the noise model, to project the distance required to reach a logical error rate of one in a trillion---known as the teraquop regime.

For the pairwise correlated model, we find similar logical error rate scaling between the correlated model and the marginalized independent model up to distance $d=15$. This indicates that surface code memory is generally robust against two-time correlated errors. We apply an exponential fit (appearing as a linear fit on a semi-logarithmic graph) of $p_L = 6.56\cdot 10^{-3}e^{-8.27 \cdot 10^{-1}d}$, for the correlated data, and a fit of $p_L = 4.03\cdot 10^{-3}e^{-8.20 \cdot 10^{-1}d}$ for the marginalized independent data. We project a teraquop distance of $27$ under both the correlated and independent noise model. A table summarizing all fits and teraquop projections throughout the paper is listed in \cref{tab:fits}.

In sharp contrast, the temporally correlated streak model exhibits slower-than-exponential suppression of the logical error rate with increasing code distance (from distance 3 to 15). Even when fitting a power-law decay as a \textit{best}-\textit{case} approximation and extrapolating, we find no realistic distance projection to achieve teraquop logical error rates. Critically, this projection assumes continued monotonic suppression of logical errors, disregarding alternative asymptotic behaviors such as saturation or a non-monotonic increase in the logical error rate. We have not definitively ruled out these more detrimental scenarios.
It is important to reiterate that the marginal per qubit error rates are well under threshold, as demonstrated by surface code performance under the marginalized independent model (\cref{fig:experiment3}). Our results show that sufficiently strong multi-time correlations in circuit-level noise can be detrimental to achieving fault tolerance with the surface code. Specifically, streaky correlations decaying quadratically or slower will, at best, result in slower-than-exponential logical error rate suppression. At distance 15, the logical per round error rate under the correlated model is measured to be $2.059\cdot 10^{-5}$, approximately 58 times the error rate under the marginalized independent model of $3.566\cdot 10^{-7}$. Under the marginalized independent model, we fit a trend of $p_L = 2.51\cdot 10^{-3}e^{-5.95 \cdot 10^{-1}d}$ and estimate a teraquop distance projection of $37$.

\Cref{fig:threshold} presents threshold plots for temporally correlated noise models compared to independent noise models with the same marginal error rates. Both the streaky correlated model and its corresponding independent model show similar \textit{apparent} thresholds in the 0.3\%–0.4\% range. The term \textit{apparent threshold} refers to the initial intersection point of logical error rate trends, but it may not necessarily correspond to a \textit{true} threshold as conventionally defined (exponential suppression as a function of code distance for surface codes). In fact, for the streaky correlated model, there are strong indications that a true threshold does not exist under a standard decoder. As we demonstrate in \cref{sec:nonmono}, further simulations reveal that for sufficiently high physical error rates, the logical error rate is a non-monotonic function of the code distance, eventually increasing after an initial decrease. While the independent model achieves exponential suppression of the logical error rate with increasing code distance below the accuracy threshold, the correlated model exhibits significantly weaker suppression. As a result, reaching teraquop-level logical error rates may require impractically large qubit overheads---or may be impossible if the logical error rate suppression follows a more pathological trend. In summary, these results so far indicate extra strategies may be required to address circuit-level noise exhibiting multi-time correlations in surface codes. In the next section, we further breakdown the cause of suboptimal surface code performance under temporally correlated noise by isolating time correlations to single classes of errors.

\section{Assessing performance impact by correlated error class}\label{sec:errorclasses}

In this section, we further investigate the detrimental impact of temporally correlated errors on the surface code by analyzing the correlations within each error class separately. We also test different correlated noise model parameters for our most pathological noise models to quantify how the strength of correlations impacts surface code performance.

Before delving into our analysis, it is helpful to reiterate the value of separating errors into different classes in the first place. While the specifics vary across hardware platforms, different error types often stem from distinct underlying physical processes. For instance, data qubit idling errors are often characterized by relaxation and dephasing errors, characterized by $T_1$ and $T_2$ times, respectively \cite{knill_randomized_2008}. In contrast, two-qubit gate errors are more likely to be tied to systemic control issues, such as frequency collisions in fixed-frequency transmon qubits \cite{rigetti_fully_2010}. Understanding how different error classes impact QEC performance can help inform platform optimization and engineering. Previous studies have shown that the logical error rate of surface code memory experiments is most sensitive to increases in Class 2 error rates, followed by Class 0 error rates, and is least affected by increases in Class 1 error rates \cite{fowler_towards_2012}.

\subsection{Surface code distance projections by correlated error class}\label{sec:errorclassesA}
\begin{figure*}[t]
    \centering
    \includegraphics[width=\textwidth]{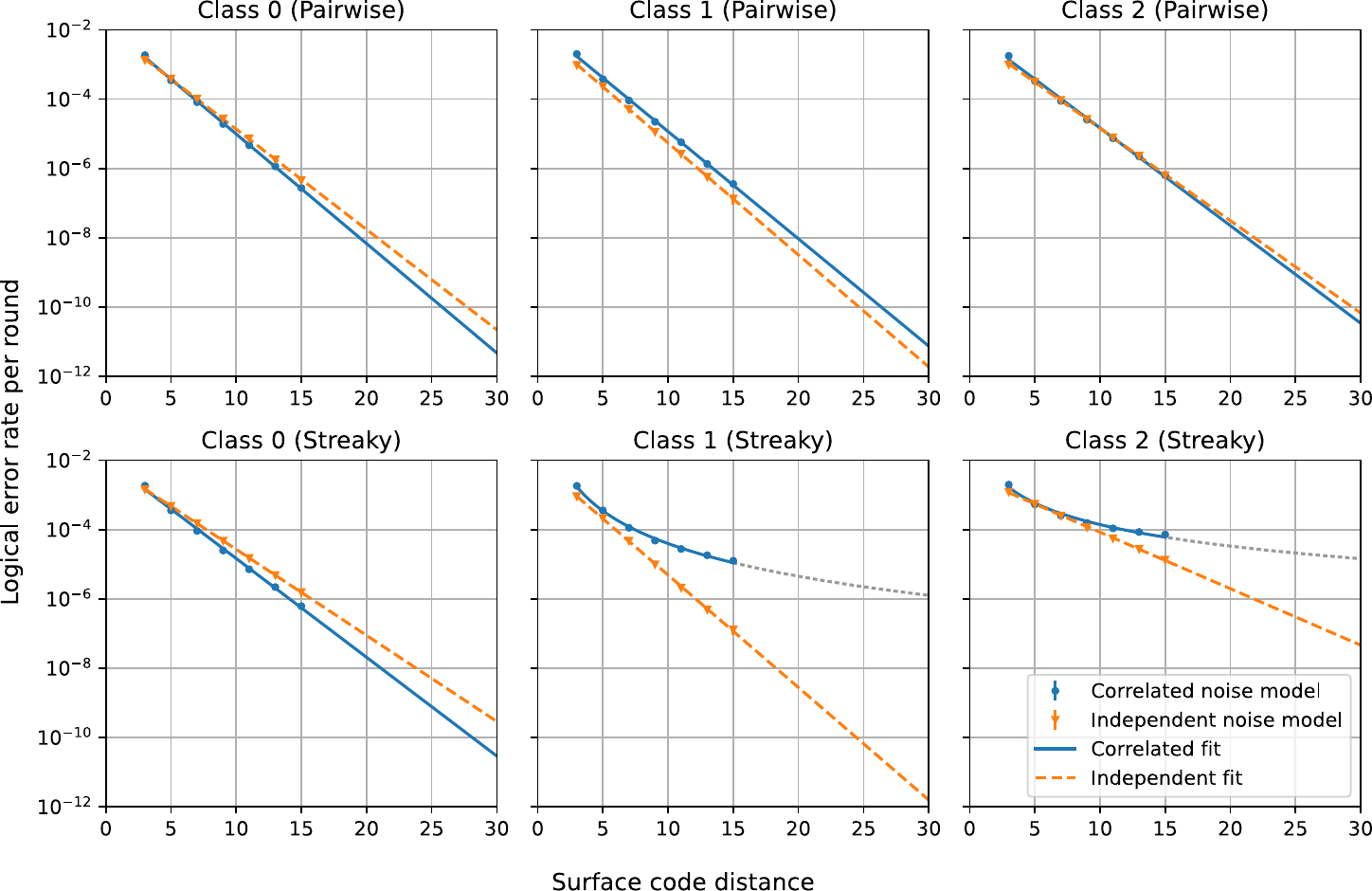}
    \caption{\label{fig:experiment1}Surface code memory logical error rate versus code distance for various noise models where correlated errors are restricted to a single error class. Error bars represent 95\% confidence intervals. Memory experiments are $2d$ rounds long, and errors are decoded using minimum weight perfect matching. All models follow the polynomial decay model $Aq/\Delta t^{n}$ with parameters $A=1, n=2, q=2\times 10^{-3}$ (i.e. quadratically decaying correlations) except for the correlated Class 2 models where $A=0.5$. Models experience independent noise beyond their correlated error class, e.g., the correlated Class 0 models additionally experience uncorrelated Class 1 and Class 2 errors. Correlated results are plotted against independent results, where the marginal error rates are fixed at each qubit at each round.}
\end{figure*}

In our first analysis, we compare the performance of the surface code under various temporally correlated noise models where correlations are restricted to a single error class. We run 10 million trials of stabilizer simulations of surface code memory experiments with $2d$ rounds, decoded using minimum weight perfect matching, inputting the same detector error model for both the correlated models and the completely independent model. All models experience independent circuit-level noise beyond their specific correlated error class. For example, the model with temporally correlated Class 0 errors will additionally experience uncorrelated Class 1 and Class 2 (but not Class 0) errors occur with probability $p=q$, the characteristic correlated error parameter. Specifically, this involves an $X$ error occurring with probability $p$ before measurement and after reset, and a $p/15$ probability each of an $IX, IY, IZ, ..., ZZ$ error occurring after each two-qubit gate. Both pairwise and streaky correlated noise models follow the polynomial decay model $Aq/t^{n}$ with parameters $A=1, n=2, q=2 \times 10^{-3}$, except for the correlated Class 2 models where $A=0.5$. We compare correlated models with independent error models, fixing the marginal error rates for each qubit at each timestep.

\Cref{fig:experiment1} plots the logical error rate per round versus surface code distance. For Class 0 correlated errors, we observe favorable scaling of logical error rate with code distance for both the pairwise and streaky model. In fact, the correlated models seem to perform slightly better than the independent models. For the correlated pairwise model, we fit an exponential trend, projecting a distance of $33$ required to reach a $10^{-12}$ per round error rate, compared to a projected distance of $35$ for the marginalized independent error model. For the correlated streaky model, we estimate a projected distance of $36$, compared to the marginalized independent model with a projected distance of $40$. From these results, surface code memory experiments appear to be robust to even our worst tested case of quadratically decaying streaky temporally correlated Class 0 errors.

The narrative shifts when we start considering correlated Class 1 errors. For the pairwise model, while we still observe favorable logical error rate scaling up to distance 15, the vertical upward shift between the correlated and marginalized independent model trends represent a small but manageable distance overhead under the correlated noise model. We measure a curve of best fit of $p_L = 1.46\cdot 10^{-2}e^{-7.13 \cdot 10^{-1}d}$ and a projected teraquop distance of $33$ for the pairwise correlated model, compared to a fit of $p_L = 9.37\cdot 10^{-3}e^{-7.45 \cdot 10^{-1}d}$, and a projected distance of $31$ for the marginalized independent model. In contrast, for the streaky correlated Class 1 model, we find that the logical error rate fails to be exponentially suppressed with increasing code distance, indicating that surface code memory is unable to cope with quadratically decaying temporally correlated streak errors. We compare this to the marginalized independent model, where we fit an exponential trend of $p_L = 8.61\cdot 10^{-3}e^{-7.48 \cdot 10^{-1}d}$ and a distance projection of $d=31$ to reach a per round error rate of $10^{-12}$. At distance 15, we measure the correlated model's logical error rate to be 97 times the independent model's logical error rate. While we still measure a reduction in logical error rate with increasing distance for the correlated error model, we find no realistic distance projection that results in error rates in the teraquop regime.

The temporally correlated Class 2 noise models exhibit similar behavior. The correlated pairwise model and its corresponding independent model display similar trends of $p_L = 9.71\cdot 10^{-3}e^{-6.49 \cdot 10^{-1}d}$ and $p_L = 6.66\cdot 10^{-3}e^{-6.14 \cdot 10^{-1}d}$, with distance projections $36$ and $37$, respectively. For the correlated Class 2 noise model, we find slower-than-exponential suppression of logical error rate with increasing distance, with no reasonable teraquop distance projection. The surface code thus does not favorably scale with less than quadratically decaying temporally correlated Class 2 streak errors. For the marginalized independent model, we measure a trend of $p_L = 3.58\cdot 10^{-3}e^{-3.75 \cdot 10^{-1}d}$, and a fault-tolerant distance projection of $59$. The relatively high distance projections in Class 2 models are attributed to the surface code’s increased sensitivity to Class 2 error rates.

Overall, these results demonstrate that while surface code memory is resilient to two-time correlated errors under certain circumstances, multi-time streak errors affecting syndrome qubits, i.e., correlated Class 1 and Class 2 errors, are highly detrimental to achieving fault tolerance with the surface code. We believe this is likely attributed to the tendency of correlated syndrome qubit errors to form timelike error strings on the syndrome graph. While an undetectable timelike error string (e.g., a syndrome measurement error at every round) does not flip the logical observable in a surface code memory experiment on its own, when combined with spacelike and spacetimelike edges arising from circuit-level noise, decoding becomes significantly more challenging, and the rate at which the decoder applies the incorrect recovery operation can rise dramatically. These findings challenge the notion the surface code logical error rate is generally least sensitive to Class 1 errors \cite{fowler_towards_2012}. Consideration of temporal correlations suggest that Class 0 noise should exhibit low physical error rates but can be highly correlated, Class 1 noise can tolerate higher error rates but must be minimally correlated, and Class 2 noise should possess both low error rates and low multi-time correlations.

We also note a link between our results and results from a recent study comparing the performance of different reset strategies for the surface code circuit \cite{geher_reset_2024}. The study found that while unconditional reset and no-reset strategies exhibited similar accuracy thresholds, the no-reset strategy would be expected to incur significant time overhead in fault-tolerant logical operations based on stability experiment simulations. This can be attributed to the idea that the no-reset strategy is a special case of the unconditional reset strategy under a temporally correlated error model. Specifically, the no-reset strategy is equivalent to an unconditional reset strategy where a syndrome measurement error is guaranteed to be followed by another syndrome reset or measurement error in the following round---that is, a locally correlated Class 1 error model.

\subsection{Varying correlated noise model parameters}\label{sec:errorclassesB}
\begin{figure*}[t]
    \centering
    \includegraphics[width=.8\textwidth]{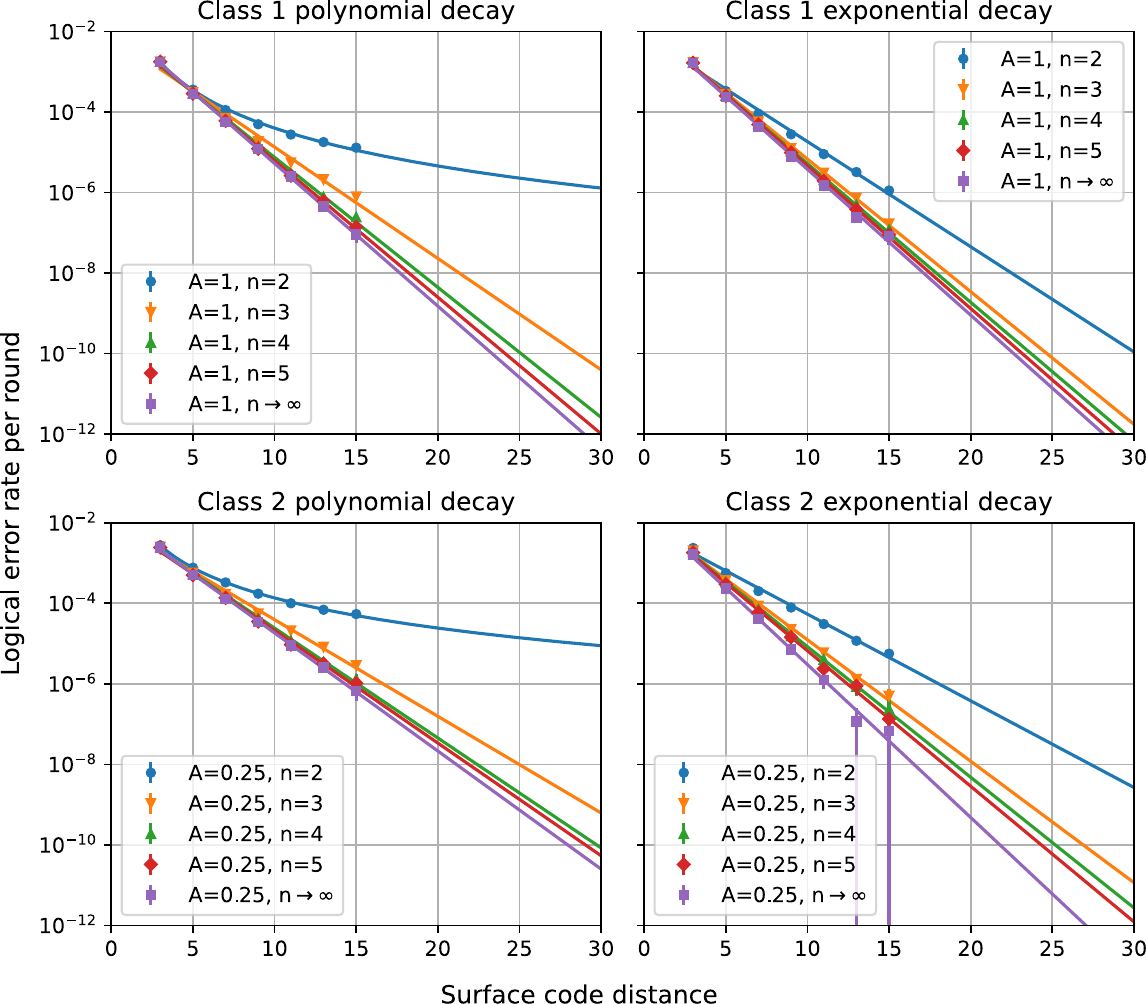}
    \caption{\label{fig:experiment2}Teraquop distance projection plots for polynomially and exponentially decaying streaky error models for Class 1 and Class 2 only errors. We test different values of $n$ in $Aq/t^n$ and $Aq/n^t$ for the polynomial and exponential model, respectively. Models with $n\to\infty$ are essentially restricted to streak lengths of only two rounds for polynomial decay models or no streaks for exponentially decay models.}
\end{figure*}
\begin{figure*}[t]
    \centering
    \includegraphics[width=\textwidth]{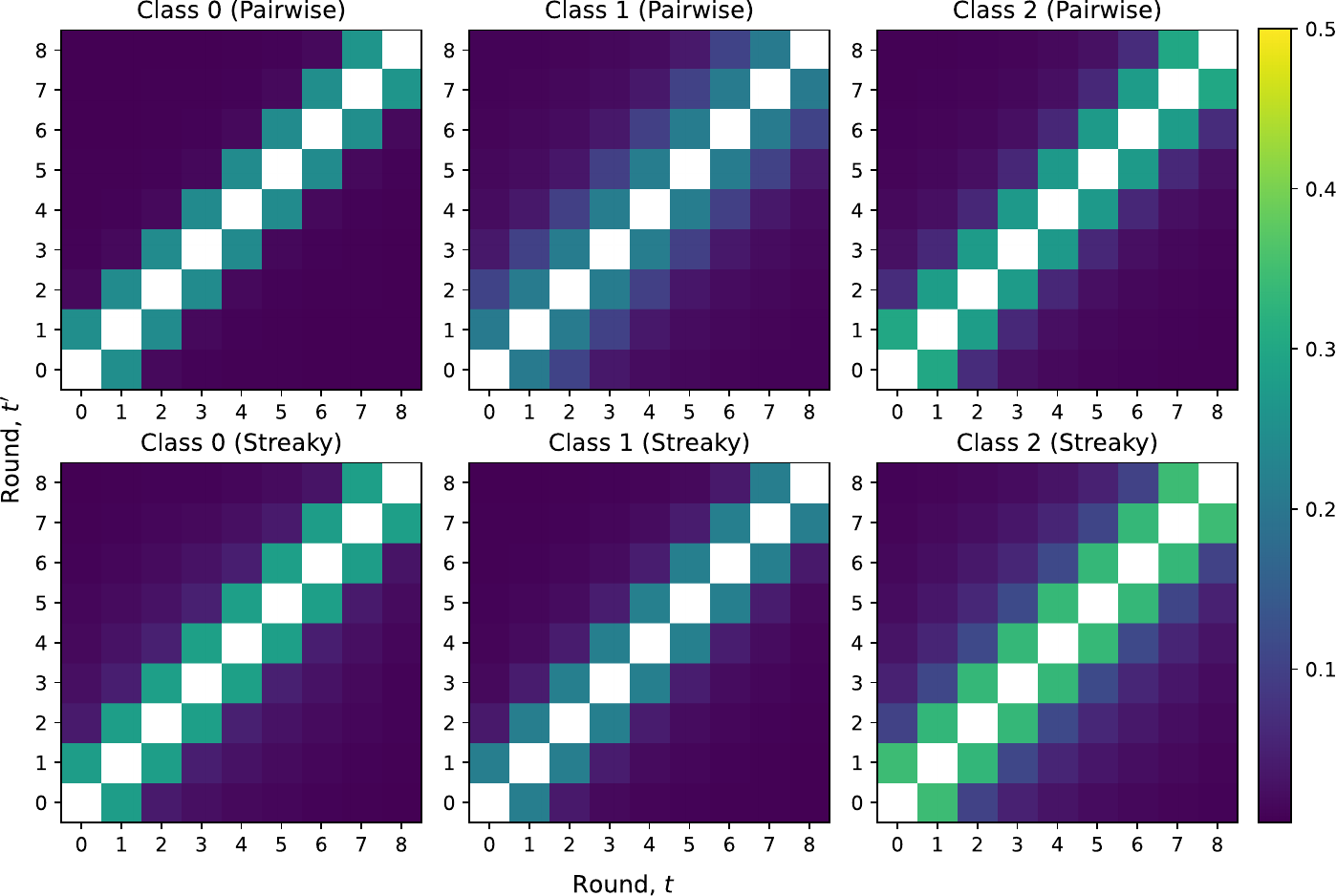}
    \caption{\label{fig:tcorrs}Detector autocorrelation matrices for a 10-round, distance 5 surface code memory experiment under various temporally correlated noise models. The initialization round is excluded since only half of the detectors are active. Self-correlations are represented by white pixels.}
\end{figure*}

\begin{figure}
    \centering
    \includegraphics[width=\columnwidth]{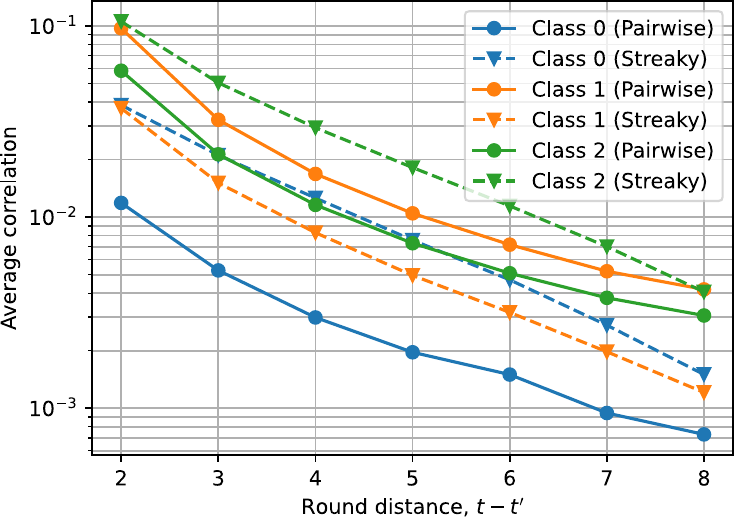}
    \caption{\label{fig:meantcorr}Mean autocorrelation $\bar{p}_{t, t'}$ versus round separation $|t-t'|$ (where $|t-t'|>1$), corresponding to \cref{fig:tcorrs}. Solid lines represent pairwise models and dashed lines represent streaky models.}
\end{figure}

In our next analysis, we investigate the relationship between correlated model parameters and surface code performance for our most pathological noise models. We run similar experiments to the previous section, with the difference that we vary the correlation decay parameter $n$ for both polynomial exponential models. \Cref{fig:experiment2} displays projection plots for temporally correlated Class 1 and Class 2 streaky error models. We test $n=2, 3, 4, 5$, and $n\to\infty$, where the latter essentially restricts streak lengths to two rounds for the polynomial decay model or no streaks at all for the exponential decay model. For Class 1 polynomially decaying correlations, we observe no reasonable teraquop distance projection for $n=2$, and distance projections for $n=3, n=4, n=5$ of $36, 32, 31$, respectively, compared to a distance projection of $29$ for $n\to\infty$. Note that although higher values of $n$ may not exhibit completely exponential error suppression behavior, high order polynomial suppression may still be sufficient to reach logical error rates in the teraquop regime with minimal distance overhead. For Class 1 exponentially decaying correlations, we observe moderate distance overhead at $n=2$, with a teraquop distance projection of $38$. We measure distance projections of $31, 30, 29, 29$ for $n=3, 4, 5$ and $n\to\infty$, respectively. We observe similar trends under the streaky correlated Class 2 models. For Class 2 polynomial decay, we observe no reasonable distance projection for $n=2$, and distance projections of $42, 38, 37, 35$ for $n=3, 4, 5$, and $n\to\infty$, respectively. For Class 2 exponential decay models, we measure distance projections of $46, 34, 32, 31, 28$, for $n=2, 3, 4, 5$, and $n\to\infty$, respectively. Overall, under Class 1 and Class 2 polynomial decay models, streaky correlations must be at least quadratically suppressed, and ideally at least cubically suppressed to achieve realistic teraquop distances, and under exponential decay models, streaky correlations are ideally exponentially suppressed with exponent $n\geq3$.

\subsection{Autocorrelation matrices under temporally correlated noise}
To assess correlations in our noise models, we measure pairwise temporal correlations between detection events by constructing autocorrelation matrices. For a Pearson correlation matrix $p_{ij}$, indices $i$ and $j$ correspond to detector coordinates, each with space ($s$) and time ($t$) components; $i=(s,t)$, $j=(s',t')$. The autocorrelation matrix $\bar{p}_{t, t'}$ which averages over all pairs $t$ and $t'$ where $s=s'$, is defined as
\begin{equation}
    \bar{p}_{t, t'} = \frac{1}{N_s}\sum_{s=s'}p_{ij},
\end{equation}
where $N_s=\sum_{s=s'}1$ is the number of detectors per round. While these matrices are useful for examining noise correlations in physical device experiments \cite{chen_exponential_2021,acharya_suppressing_2023,miao_overcoming_2023}, they are limited by their inability to measure multi-time correlations. \Cref{fig:tcorrs} shows autocorrelation matrices $\bar{p}_{t, t'}$ for a 10-round, distance 5 surface code memory experiment under each of the correlated noise models shown in \cref{fig:experiment1}. Self-correlations are colored white. \cref{fig:meantcorr} plots the mean autocorrelation versus separation in rounds $|t-t'|$, excluding $|t-t'|=1$. 

Each correlated noise model exhibits non-zero correlations up to a time separation of $|t-t'|=8$, where the Class 0 pairwise model shows the weakest pairwise correlations, and the Class 2 streaky model the strongest. Interestingly, the streaky versions of the Class 0 and Class 2 models demonstrate stronger pairwise correlations than their corresponding pairwise models, whereas the opposite trend is observed for the Class 1 models. The degree of pairwise correlations in detection events under a given noise model does not directly correspond to the severity of its impact on surface code performance. This is likely because pairwise correlations do not distinguish between continuous, stringlike errors and disjoint localized errors, even though the former is potentially more damaging than the latter. For example, in the Class 0 pairwise model, pairwise detector correlations arise due to correlated pairs of spacelike errors on the same data qubit, separated a number of rounds apart. In contrast, in the Class 1 streaky model, correlations correspond to ends of error streaks on syndrome qubits.

Overall, these results indicate that alternative characterization techniques, that can distinguish multi-time correlations, may be required to better differentiate benign and catastrophic multi-time correlated error structures.

%\section{Analysis}\label{sec:analysis}
\section{Discussion}\label{sec:discussion}
In this section, we discuss our perspective on how certain temporally correlated error structures are more detrimental than others. We discuss our results in relation to recent experimental evidence of a noise floor in superconducting QEC experiments \cite{acharya_quantum_2024} attributed to rare correlated events. We also briefly discuss strategies for mitigating pathological temporally correlated error structures, such as tailoring decoding algorithms for specific structures.

\subsection{Temporally correlated noise-induced logical errors}
\begin{figure}
    \centering
    \subfloat[\label{fig:topological1}Independent]
    {\includegraphics[width=.32\columnwidth]{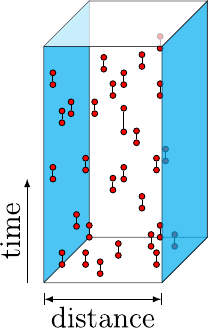}\hspace{2mm}}
    \subfloat[\label{fig:topological2}Correlated]
    {\includegraphics[width=.32\columnwidth]{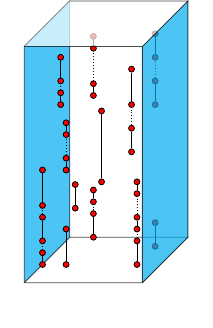}}
    \subfloat[\label{fig:topological3}Logical Error]
    {\includegraphics[width=.32\columnwidth]{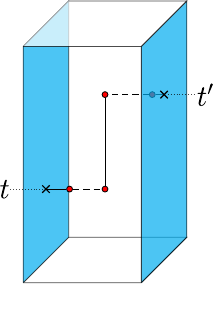}}
    \caption{\label{fig:topological}Topological perspective on temporally correlated error structures. \protect\subref{fig:topological1} portrays timelike errors in surface code memory under an independent noise model. Timelike strings of predominantly length-one are randomly distributed throughout the bulk. \protect\subref{fig:topological2} portrays timelike errors under a temporally correlated Class 1 streaky model. Timelike errors coalesce into larger chains of errors. \protect\subref{fig:topological3} shows an example of a logical error due to one long, temporal stringlike error, with vertices at $t$ and $t'$, and two spatial strings.}
\end{figure}

\Cref{fig:topological} presents a topological illustration of correlated and uncorrelated timelike (syndrome) errors in surface code memory. In the independent model, shown in \cref{fig:topological1}, timelike strings, predominantly of length one, are randomly distributed throughout the bulk. Conversely, in the correlated streaky model (but not the pairwise model), shown in \cref{fig:topological2}, timelike strings coalesce into larger temporal chains. These temporal chains are not necessarily single continuous stringlike errors but are generally composed of timelike strings with lengths greater than one. Undetectable logical errors correspond to stringlike errors that connect and terminate at opposite and like boundaries. While a purely temporal stringlike error does not cause a logical error in surface code memory by itself, they can interact with spatial string errors in a manner that increases the likelihood of a logical error. Note that in some surface code operations, such as in stability experiments or lattice surgery \cite{gidney_stability_2022}, temporal stringlike errors \textit{can} directly cause a logical error.

\Cref{fig:topological3} illustrates one such logical error involving a temporal stringlike error. Here, solid lines represent chains of physical errors, consisting of a single temporal string spanning $t$ and $t'$, and a spatial string at both $t$ and $t'$. Dashed lines represent the minimum-weight error for the given syndrome, in this case, a pair of spatial strings at $t$ and $t'$. The temporal stringlike error and spatial errors thus combine to apply an unwanted logical operation. In essence, temporally correlated errors on syndrome qubits forming temporal strings can significantly increase the likelihood that errors on data qubits (spatial strings) will ``accumulate” between rounds, even if they are separated far apart. This interaction can cause a logical error that would not have occurred if errors between rounds were strongly temporally isolated, as is the case under temporally uncorrelated noise. This also likely explains why surface code memory is seemingly resilient to multi-time correlated errors on data qubits, as error streaks on data qubits (even those producing measurable temporal correlations in syndrome measurement outcomes) do not form temporal stringlike errors in the same manner that error streaks on syndrome qubits do. %It is expected correlated errors become more important as physical error rates decrease \cite{harper_learning_2023,acharya_suppressing_2023}.

We remark a connection between our work and recent experiments on surface code memory using superconducting qubits \cite{acharya_quantum_2024}. The study observed an apparent logical-error-per-cycle floor of $10^{-10}$ for high-distance repetition codes. The noise floor was attributed to rare correlated error events. One type of event involved a sharp increase in simultaneous detection events, while the other involved a single noisy detector event over thousands of rounds. The latter type of correlated error event closely resembles the multi-time correlated streaky errors on syndrome qubits. This underscores the importance of understanding how temporally correlated error structures interact with topological QEC protocols and mitigating their negative effects on code performance. This is especially crucial when considering non-memory surface code operations, such as lattice surgery, which are potentially more sensitive to temporal stringlike errors than surface code memory.

\subsection{Potential mitigation strategies}
Structure in noise can impact the efficacy of QEC protocols. Conversely, knowledge of this structure can be leveraged to enhance error correction capabilities potentially beyond cases assumed under i.i.d. models \cite{tuckett_ultrahigh_2018,tuckett_tailoring_2019,tuckett_fault-tolerant_2020}. Temporally correlated errors are no exception in this regard. We briefly discuss potential strategies for exploiting temporal structure in noise. 

The efficacy of different approaches will depend on the specific characteristics of the temporally correlated noise model. For instance, if temporal stringlike errors of a certain weight dominate the error probability distribution, it may be possible to improve logical error rates by designing specialized decoding algorithms that assign higher weights to long-range edges of certain lengths, similar to the protocol proposed in a previous study \cite{nickerson_analysing_2019}. On the other hand, if temporal error chains are sufficiently rare, long, and detectable, as observed in recent experiments on superconducting devices \cite{acharya_quantum_2024}, one can employ similar techniques as in drift-aware decoders, where edge weights are dynamically updated based on recent syndrome history \cite{wang_dgr_2024}. Alternatively, code deformation can be used to render defective data or syndrome qubits in a surface code patch permanently or temporarily inactive, such as in fabrication defects or cosmic ray mitigation protocols \cite{auger_fault-tolerance_2017,strikis_quantum_2023,siegel_adaptive_2023}. However, correlated noise that does not fit neatly into either of these categories, such as the models explored in this paper, may require the development of new mitigation strategies that better account for general temporal correlation structures.

Another promising avenue for mitigating temporal stringlike errors is employing \textit{walking} surface code circuits \cite{mcewen_relaxing_2023}. Unlike the standard surface code, which has fixed stabilizers, walking surface codes move around stabilizers on the underlying qubit grid while incurring minimal resource overheads. These dynamic codes have the relevant property that data qubits and syndrome qubits exchange roles between error correction cycles. This property can be useful not only for leakage mitigation but also for breaking up harmful temporal stringlike errors. Although there are numerous existing protocols that seek to mitigate certain correlated error processes or structures, as QEC experiments continue to scale up in size and complexity, it is likely new protocols will need to be developed to address the complex zoo of spatio-temporal correlations in noise affecting quantum devices.

\subsection{Conclusion}
We have explored the impact of temporally correlated error structures, based off circuit-level noise, on the performance of surface code memory. Our analysis has found that while fault tolerance is realistically achievable under arbitrary long-range two-time correlated errors, multi-time ``streaky" correlated errors involving syndrome qubits can significantly degrade logical error rate scaling. Specifically, extrapolating simulation data on logical error rates, we find that quadratically decaying streaky correlations result in, at best, slower-than-exponential suppression, with no feasible distance projection for achieving teraquop error rates.

Our results underscore the importance of understanding and mitigating correlated noise in QEC protocols. We demonstrated that surface code memory is resilient to temporally correlated errors on data qubits but highly sensitive to similar errors on syndrome qubits. We attribute this sensitivity to the formation of temporal stringlike errors, which complicate the decoding process and increase the likelihood of logical errors. Potential strategies for mitigating these effects include tailored decoding algorithms, dynamic code deformation techniques, and employing walking surface code circuits. Further research is needed to characterize a broader class of non-Markovian spatio-temporal correlated error structures and develop effective mitigation protocols, to further elucidate pathways towards fault-tolerant quantum computing.

\begin{acknowledgments}
\noindent 
All authors, except the first, are listed in alphabetical order. J. K. is supported by an Australian Government Research Training Program Scholarship and a CSIRO Future Science Platform Scholarship. K. M. is grateful for the support of the Australian Research Council via Discovery Projects DP210100597 and DP220101793.
\end{acknowledgments}
\bibliography{references}

\clearpage
\onecolumngrid
\appendix
\renewcommand\thefigure{\thesection\arabic{figure}}
\section{Error marginalization}\label{sec:marginalization}
\setcounter{figure}{0} 

In this appendix, we outline our method for obtaining marginal error probabilities from a temporally correlated noise model. We then provide a concrete, worked example of the marginalization process for a pairwise time-correlated correlated error model.

For convenience, we begin by reparametrizing the error channels given in \cref{eqn:channels} as follows.
\begin{subequations}\label{channels-app}
    \begin{align}
    &\mathcal{E}^1_X(p'{,\rho}) = (1-p')\rho + \frac{p'}{2}\left(\rho + X\rho X\right)\label{eqn:channelX_app} \\
    &\mathcal{E}^1_D(p'{,\rho}) = (1-p')\rho + \frac{p'}{4}(\rho + X\rho X + Y\rho Y + Z\rho Z)\label{eqn:channelD1_app} \\
    &\mathcal{E}^2_D(p'{,\rho}) = (1-p')\rho + \frac{p'}{16}\sum_{P\in\{I,X,Y,Z\}^{\otimes 2}}P\rho P\label{eqn:channelD2_app}.
    \end{align}
\end{subequations}
Each of these channels is equivalent to the corresponding channel in \cref{eqn:channels} under the transformation $p' = C_{\mathcal{E}} p$, where $C_\mathcal{E}$ is a constant that depends on the error channel $\mathcal{E}$. Specifically, 
\begin{align}
    C_\mathcal{E}=
    \begin{cases}
        2, & \mathcal{E} = \mathcal{E}^1_X\\
        \frac{4}{3}, & \mathcal{E} = \mathcal{E}^1_D\\
        \frac{16}{15}, & \mathcal{E} = \mathcal{E}^2_D\\
    \end{cases}.
\end{align}
This moves away from the standard Stim uses (\cref{eqn:channels}) to the equivalent interpretation in which a channel $\mathcal{E}(p',\rho)$ maximally ``mixes" a state $\rho$ with probability $p'$, where the maximally mixing channel $\mathcal{M}$ has the convenient property $(\mathcal{M}\circ\mathcal{M})[\rho]=\mathcal{M}[\rho]$. Consequently, we define $p'(s,t)$ as the probability that qubit $s$ at time $t$ is maximally mixed by the noise channel. To compute the probability $p(s,t)$, we first compute $p'(s,t)$ and then apply the conversion using the factor $C_\mathcal{E}$, where $\mathcal{E}$ implicitly depends on both $s$ and $t$ (i.e., what gate immediately precedes the error channel).

We reparametrize the channels in this way to enable the use of the following formula for calculating the probability that an error occurs at any given space-time coordinate in the experiment. If qubit $s$ at time $t$ can experience $k$ possible errors, each of which maximally mixes the qubit and occurs independently with probability $p_i'$, for $i \in {1, \dots, k}$, then the probability that no mixing occurs is given by:
\begin{align}\label{eqn:no-events}
    1-p'(s,t) = \prod_{i=1}^k (1 - p_i').
\end{align}

This formula is based on the property that applying two maximally mixing channels has the same effect as applying a single maximally mixing channel: $(\mathcal{M}\circ\mathcal{M})[\rho] = \mathcal{M}[\rho]$. Thus, the probability that the qubit does not become mixed is simply the probability that none of the $k$ ``mixing" errors occur. This holds for single Pauli flip errors as well, where $p'(s,t)$ is twice the probability that qubit $s$ at time $t$ experiences a Pauli flip.

Define $\mathcal{T}$ as the set of all error events that can occur during the experiment. Each event $E \in \mathcal{T}$ occurs independently with probability $\mathrm{Pr}(E)$. This event does not necessarily correspond to a single error channel but rather to a set of possible error channels $\{X_1^{E}, \dots, X^{E}_m\}$,  with one channel selected with probability $\mathbb{P}\left( X_i^{E} \mid E \right)$. 

For the models considered in this paper, the set of all error events affecting qubit $s$ is denoted as $\mathcal{T}_s := \{E_{s,i,j}~|~ 1 \leq i < j \leq N\}$, where $N$ is the number of rounds in the experiment; it follows that $\mathcal{T}$ is simply the (disjoint) union of $\mathcal{T}_s$ for all $s$. In the pairwise model, the set of errors corresponding to $E_{s,i,j}$ is the set of all Pauli strings on qubit $s$ at times $i$ and $j$, i.e., 
\begin{align}
    \{I_i, X_i, Y_i, Z_i\}\otimes \{I_j, X_j, Y_j, Z_j\}.
\end{align}
In the streaky model, the set of errors is given by
\begin{align}
    \bigotimes_{k=i}^{j} \{I_k, X_k, Y_k, Z_k\}.
\end{align}
In both models, the corresponding distribution $\mathbb{P}$ over these sets is simply the uniform distribution. That is, if error event $E_{s,i,j}$ occurs, qubit $s$ experiences a uniformly random Pauli error (including the possibility of the identity) at times $i$ and $j$ (and also times $i \leq t \leq j$ in the case of the streaky model).
Then, for the pairwise model, it follows from \cref{eqn:no-events} that $p'(s,t)$, the probability that qubit $s$ is maximally mixed at time $t$, satisfies
\begin{align}\label{eqn:marginalpairwise}
    1-p'(s,t) = \prod_{1 \leq i < j \leq N} \left( 1 - \mathrm{Pr}\left(E_{s, i, j} \right) \mathbbm{1}\{t \in \{i, j\} \} \right),
\end{align}
where $\mathbbm{1}\{t \in \{i, j\} \}$ is $1$ if $t\in \{i,j\}$ and $0$ otherwise. In the streaky model, we similarly obtain
\begin{align}\label{eqn:marginalstreaky}
    1-p'(s,t) = \prod_{1 \leq i < j \leq N} \left( 1 - \mathrm{Pr}\left(E_{s, i, j} \right) \mathbbm{1}\{t \in [i, j] \} \right).
\end{align}
In the example below we write $u_{ij}(t) = \mathrm{Pr}\left(E_{s, i, j} \right) \mathbbm{1}\{t \in \{i, j\} \}$ for a fixed $s$. Thus, $p(s,t)$ is computed by 
\begin{align}\label{eqn:marginalfinal}
    p(s,t) = \frac{1}{C_\mathcal{E}} \left[1 - \prod_{1 \leq i < j \leq N} \left( 1 - u_{ij}(t) \right) \right].
\end{align}

\subsection*{Example: Pairwise correlated depolarizing errors}

Consider a pairwise time-correlated depolarizing error model on a single qubit $s$ over $N$ timesteps. We wish to write the marginalized probability distribution $p(s,t)=p(t)$ for $t\in\{1,2,\dots,N\}$ (suppressing $s$ in the notation as we focus on just a single qubit), where $p(t)$ is the probability the qubit experiencing a uniformly random non-identity Pauli error at timestep $t$ --- i.e., $\mathcal{E}(t)=\mathcal{E}^1_D(p(t))$. Define the probability that 1 of 15 equally likely pairs of errors $IX, IY, IZ,\dots, ZZ$ occur at timesteps $i$ and $j$ to be $Aq/|i-j|^n$. This matches the definition of the pairwise error model given in Section~\ref{sec:noisemodeldetails}. Then the probability the qubit at timestep $t$ becomes maximally mixed (equal chance $I, X, Y, Z$) due to each error event $E_{ij}|t\in\{i,j\}$ is
\begin{equation}\label{eqn:examplepairwisedepol}
    u_{ij}(t) = \frac{16}{15}\frac{Aq}{|i-j|^n}.
\end{equation}
The factor of $16/15$ is due to how we defined our error model in the main text, where we followed the convention Stim uses to define Pauli error channels. Thus, inputting \cref{eqn:examplepairwisedepol} into \cref{eqn:marginalfinal} and taking $C_{\mathcal{E}}=\frac{4}{3}$, we obtain
\begin{equation}
    p(t) = \frac{3}{4} \left[1 - \prod_{1 \leq i < j \leq N} \left( 1 - \frac{16Aq}{15|i-j|^n} \right) \right].
\end{equation}

\section{Non-monotonicity of logical error rate scaling under detrimental noise models}\label{sec:nonmono}
\setcounter{figure}{0} 

In the main text, our analysis of the streaky correlated noise model fitted the logical error rate to a power-law decay, as shown in the bottom panel of \cref{fig:experiment3}. However, this fit does not preclude the possibility of more pathological behavior. Small but systematic deviations between the data and the power-law trend suggest that the observed scaling may not persist at larger code distances. In particular, the logical error rate could exhibit non-monotonic behavior---initially decreasing with code distance, but eventually saturating or even increasing at larger distances.

\begin{figure}
    \centering
    \includegraphics[width=0.6\columnwidth]{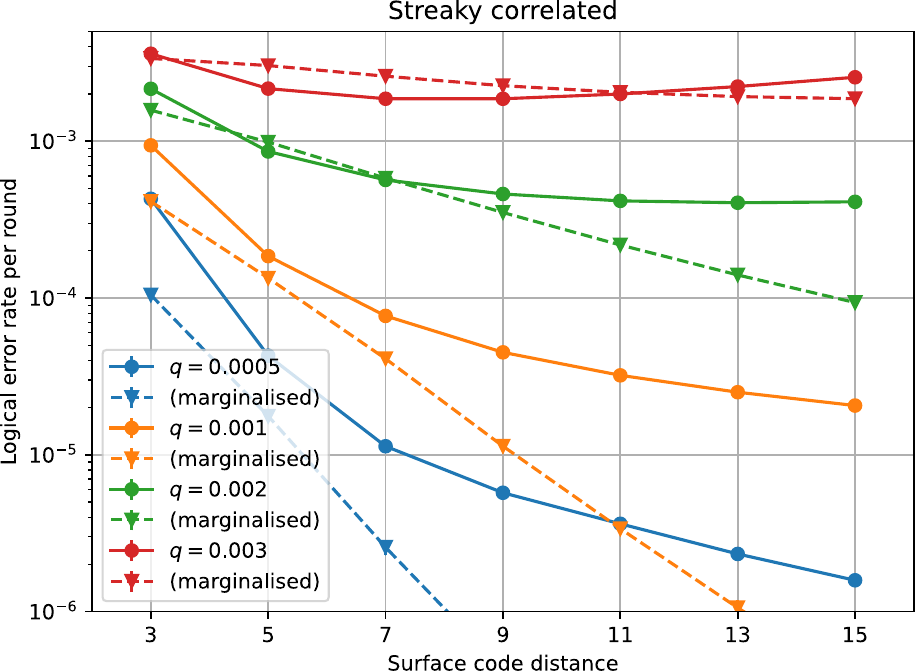}
    \caption{\label{fig:experiment7}Logical error rate per round versus surface code distance under the streaky correlated model. Data is shown for several correlated error parameters. The data for $q=1\times 10^{-3}$ corresponds to the data shown in the bottom panel of \cref{fig:experiment3}. For higher values of $q$, the logical error rate scaling is non-monotonic, reaching a minimum before increasing at larger code distances. We also display data for the corresponding marginalized error models for parameters $q=0.002$ and $q=0.003$. Of note is the marginalized model data for $q=0.002$ which indicates exponential suppression of logical error rate for uncorrelated errors in contrast to the non-monotonic behavior of the corresponding correlated model.}
\end{figure}

To further probe this behavior, we performed additional simulations of the streaky correlated model while varying the characteristic correlated error probability $q$. The results, shown in \cref{fig:experiment7}, reveal instances of non-monotonic logical error rate scaling. For $q=2\times10^{-3}$, the logical error rate at $d=11$ exceeds that at $d=13$; similarly, for $q=3\times10^{-3}$, the logical error rate at $d=11$ exceeds that at $d=9$, further increasing with code distance. It is worth noting that for $q=3\times10^{-3}$, the marginal physical error rate approaches the surface code threshold, which also leads to degraded performance in the corresponding marginalized independent model---though this degradation remains monotonic.

These results provide evidence that, for sufficiently high physical error rates and code distances, the logical error rate under streaky correlated noise can exhibit non-monotonic behavior---even when the corresponding marginalized independent noise model continues to show the expected monotonic suppression. Whether such non-monotonic scaling persists across all values of the characteristic correlated parameter $q$ remains an open question and may require rigorous analytical treatment to resolve. Regarding the existence of a threshold, we emphasize that even under monotonic decay, the absence of exponential suppression is sufficient to disqualify a \textit{true} threshold as conventionally defined \cite{fowler_towards_2012}. This is why we refer to an ``apparent threshold" in the main text. Altogether, these findings reinforce our central conclusion: certain temporally correlated noise structures can fundamentally compromise the fault-tolerance properties of surface codes when using a standard decoder.

\begin{table*}
\caption{\label{tab:fits}Table of fits and teraquop distance projections for various surface code memory experiments under temporally correlated noise models. The error class describes which gates on which qubits are affected by correlated noise, the structure describes whether correlations are two-time (pairwise) or multi-time (streaky), and the decay function and parameters define the strength of correlations.}
\begin{ruledtabular}
\begin{tabular}{l c c c c c}
\textbf{Error Class} & \textbf{Structure} & \textbf{Decay Function} & \textbf{} \textbf{Parameters}($A, q, n$) & \textbf{Fit} & \textbf{Teraquop Distance}\\ \hline
\multicolumn{6}{c}{\cref{sec:fullcircuitlevel}, \cref{fig:experiment3}} \\ \hline
Circuit-level & Pairwise & Polynomial & $1, 10^{-3}, 2$ & $6.56\cdot 10^{-3}e^{-8.27 \cdot 10^{-1}d}$ & 27\\
(Marginalized) &&&& $4.03\cdot 10^{-3}e^{-8.20 \cdot 10^{-1}d}$ & 27\\
Circuit-level & Streaky & Polynomial & $1, 10^{-3}, 2$ & $9.51\cdot 10^{-3}d^{-2.35}$  & -*\\
(Marginalized) &&&& $2.51\cdot 10^{-3}e^{-5.95 \cdot 10^{-1}d}$ & 37\\ \hline

\multicolumn{6}{c}{\cref{sec:errorclassesA}, \cref{fig:experiment1}} \\ \hline
Class 0 & Pairwise & Polynomial & $1, 2\times 10^{-3}, 2$ & $1.43\cdot 10^{-2}e^{-7.28\cdot 10^{-1}d}$ & 33\\
(Marginalized) &&&& $1.07\cdot 10^{-2}e^{-6.67\cdot 10^{-1}d}$ & 35\\

Class 0 & Streaky & Polynomial & $1, 2\times 10^{-3}, 2$ & $1.04\cdot 10^{-2}e^{-6.57 \cdot 10^{-1}d}$ & 36\\
(Marginalized) &&&& $8.23\cdot 10^{-3}e^{-5.72\cdot 10^{-1}d}$ & 40\\

Class 1 & Pairwise & Polynomial & $1, 2\times 10^{-3}, 2$ & $1.46\cdot 10^{-2}e^{-7.13 \cdot 10^{-1}d}$ & 33\\
(Marginalized) &&&& $9.40\cdot 10^{-3}e^{-7.45 \cdot 10^{-1}d}$ & 31\\

Class 1 & Streaky & Polynomial & $1, 2\times 10^{-3}, 2$ & $5.38\cdot 10^{-2}d^{-3.13}$ & -*\\
(Marginalized) &&&& $8.61\cdot 10^{-3}e^{-7.48\cdot 10^{-1}d}$ & 31\\

Class 2 & Pairwise & Polynomial & $0.5, 2\times 10^{-3}, 2$ & $9.71\cdot 10^{-3}e^{-6.49\cdot 10^{-1}d}$ & 36\\
(Marginalized) &&&& $6.66\cdot 10^{-3}e^{-6.14\cdot 10^{-1}d}$ & 37\\

Class 2 & Streaky & Polynomial & $0.5, 2\times 10^{-3}, 2$ & $1.62\cdot 10^{-2}d^{-2.07}$ & -*\\
(Marginalized) &&&& $3.58\cdot 10^{-3}e^{-3.75 \cdot 10^{-1}d}$ & 59\\ \hline

\multicolumn{6}{c}{\cref{sec:errorclassesB}, \cref{fig:experiment2}} \\ \hline
Class 1 & Streaky & Polynomial & $1, 2\times 10^{-3}, 3$ & $7.85 \cdot 10^{- 3} e^{-6.37 \cdot 10^{- 1}d}$ & 36\\
Class 1 & Streaky & Polynomial & $1, 2\times 10^{-3}, 4$ & $1.27 \cdot 10^{- 2} e^{-7.43 \cdot 10^{- 1}d}$ & 32\\
Class 1 & Streaky & Polynomial & $1, 2\times 10^{-3}, 5$ & $1.53 \cdot 10^{- 2} e^{-7.81 \cdot 10^{- 1}d}$ & 31\\
Class 1 & Streaky & Polynomial & $1, 2\times 10^{-3}, \infty$ & $1.85 \cdot 10^{- 2} e^{-8.16 \cdot 10^{- 1}d}$ & 29\\
Class 1 & Streaky & Exponential & $1, 2\times 10^{-3}, 2$ & $7.54 \cdot 10^{- 3} e^{-6.01 \cdot 10^{- 1}d}$ & 38\\
Class 1 & Streaky & Exponential & $1, 10^{-3}, 3$ & $1.33 \cdot 10^{- 2} e^{-7.58 \cdot 10^{- 1}d}$ & 31\\
Class 1 & Streaky & Exponential & $1, 10^{-3}, 4$ & $1.46 \cdot 10^{- 2} e^{-7.93 \cdot 10^{- 1}d}$ & 30\\
Class 1 & Streaky & Exponential & $1, 10^{-3}, 5$ & $1.62 \cdot 10^{- 2} e^{-8.16 \cdot 10^{- 1}d}$ & 29\\
Class 1 & Streaky & Exponential & $1, 10^{-3}, \infty$ & $1.59 \cdot 10^{- 2} e^{-8.34 \cdot 10^{- 1}d}$ & 29\\

Class 2 & Streaky & Polynomial & $0.25, 10^{-3}, 3$ & $9.83 \cdot 10^{- 3} e^{-5.53 \cdot 10^{- 1}d}$ & 42\\
Class 2 & Streaky & Polynomial & $0.25, 10^{-3}, 4$ & $1.3 \cdot 10^{- 2} e^{-6.28 \cdot 10^{- 1}d}$ & 38\\
Class 2 & Streaky & Polynomial & $0.25, 10^{-3}, 5$ & $1.34 \cdot 10^{- 2} e^{-6.44 \cdot 10^{- 1}d}$ & 37\\
Class 2 & Streaky & Polynomial & $0.25, 10^{-3}, \infty$ & $1.57 \cdot 10^{- 2} e^{-6.75 \cdot 10^{- 1}d}$ & 35\\

Class 2 & Streaky & Exponential & $0.25, 10^{-3}, 2$ & $7.64 \cdot 10^{- 3} e^{-4.96 \cdot 10^{- 1}d}$ & 46\\
Class 2 & Streaky & Exponential & $0.25, 10^{-3}, 3$ & $1.28 \cdot 10^{- 2} e^{-6.94 \cdot 10^{- 1}d}$ & 34\\
Class 2 & Streaky & Exponential & $0.25, 10^{-3}, 4$ & $1.46 \cdot 10^{- 2} e^{-7.47 \cdot 10^{- 1}d}$ & 32\\
Class 2 & Streaky & Exponential & $0.25, 10^{-3}, 5$ & $1.54 \cdot 10^{- 2} e^{-7.75 \cdot 10^{- 1}d}$ & 31\\
Class 2 & Streaky & Exponential & $0.25, 10^{-3}, \infty$ & $1.89 \cdot 10^{- 2} e^{-8.75 \cdot 10^{- 1}d}$ & 28\\
\end{tabular}
\end{ruledtabular}
Polynomial: $Aq/\Delta t^n$, Exponential: $Aq/n^{\Delta t}$, *No realistic teraquop projection
\end{table*}

\end{document}